\documentclass[prd,aps,twocolumn,preprintnumbers, showpacs, nofootinbib,superscriptaddress,notitlepage]{revtex4-1}

\usepackage{amssymb, amsthm, amsmath} 
\usepackage{graphicx}   
\usepackage{color}      
\usepackage{hyperref}

\begin{document}

\preprint{MSUHEP-19-001}
\title{Lattice-QCD Determination of the Hyperon Axial Couplings in the Continuum Limit}

\author{Aditya Savanur}
\affiliation{Department of Physics and Astronomy, Michigan State University, East Lansing, MI 48824, USA}

\author{Huey-Wen Lin}
\email{hwlin@pa.msu.edu}
\affiliation{Department of Physics and Astronomy, Michigan State University, East Lansing, MI 48824, USA}
\affiliation{Department of Computational Mathematics, Michigan State University, East Lansing, MI 48824, USA}


\begin{abstract}
We present the first continuum extrapolation of the hyperon octet axial couplings ($g_{\Sigma \Sigma}$ and $g_{\Xi \Xi}$) from $N_f=2+1+1$ lattice QCD. 
These couplings are important parameters in the low-energy effective field theory description of the octet baryons and fundamental to the nonleptonic decays of hyperons 
and to hyperon-hyperon and hyperon-nucleon scattering with application to neutron stars.
We use clover lattice fermion action for the valence quarks with sea quarks coming from configurations of $N_f=2+1+1$ highly improved staggered quarks (HISQ) generated by MILC Collaboration. 
Our work includes the first calculation of $g_{\Sigma \Sigma}$ and $g_{\Xi \Xi}$ directly at the physical pion mass on the lattice, and a full account of systematic uncertainty, including excited-state contamination, finite-volume effects and continuum extrapolation, all addressed for the first time. 
We find the continuum-limit hyperon coupling constants to be
$g_{\Sigma \Sigma}=0.4455(55)_\text{stat}(65)_\text{sys}$ and
$g_{\Xi \Xi}     =-0.2703(47)_\text{stat}(13)_\text{sys}$,
which correspond to low-energy constants of 
$D = 0.708(10)_\text{stat}(6)_\text{sys}$ and
$F = 0.438(7)_\text{stat}(6)_\text{sys}$.
The corresponding SU(3) symmetry breaking is 9\% which is about a factor of 2 smaller than the earlier lattice estimate. 
\end{abstract}

\maketitle

\section{Introduction}

The octet-baryon axial couplings ($g_{\Sigma \Sigma}$, $g_{\Xi \Xi}$, and $g_A$) are important quantities for studying hadron structure in QCD. Specifically, the hyperon couplings are important in the effective field theory of octet baryons~\cite{Savage:1996zd}, because they enter the expansions of all quantities in chiral perturbation theory. 
In addition, the coupling constants also appear in the calculations of hyperon nonleptonic decays~\cite{Cabibbo:2003cu} and of hyperon-hyperon and hyperon-nucleon scattering matrix elements~\cite{Beane:2003yx}. 
They are also useful for calculating equations of state and other properties of nuclear matter in neutron stars~\cite{Lattimer:2006xb,Weissenborn:2011ut}. 
Studying these couplings also allows us to explore the extent of the symmetry breaking of SU(3) flavor. 
SU(3) symmetry has been widely studied~\cite{Savage:1996zd,Dai:1995zg} in the hyperon hadronic matrix elements and this symmetry is used in many applications where strange data is limited. 
For example, the global analysis of the polarized parton distribution function (PDF) has commonly used this assumption for extracting individual quark flavor PDFs~\cite{Lin:2017snn}; 
knowing to what extent of this symmetry holds will help us quantify the systematic uncertainty introduced by the use of this assumption in the polarized PDF~\cite{Lin:2017snn}. 
However, experimentally it is much harder to determine the hyperon couplings than those in the nucleon case, since the hyperons weak decay in nature quickly. 
Lattice-QCD (LQCD) calculations can provide more stringent direct and reliable calculations of these couplings. 

Lattice QCD is an ideal theoretical tool to study the parton structure of hadrons, starting from quark and gluon degrees of freedom.  Progress has long been limited by computational resources, but recently advances in both algorithms and a worldwide investment in pursuing exascale computing has led to exciting progress in LQCD calculations. Take the nucleon tensor charge for example. Experimentally, one gets the tensor charges by taking the zeroth moment of the transversity distribution; however, the transversity distribution is poorly known and such a determination is not very accurate. 
On the lattice side, there are a number of calculations of $g_T$~\cite{Gupta:2018qil,Lin:2018qky,Bhattacharya:2016zcn,Bhattacharya:2015wna,Green:2012ej,Aoki:2010xg,Abdel-Rehim:2015owa,Bali:2014nma,Yamazaki:2008py}; some of them are done with more than one ensemble at physical pion mass with high-statistics calculations (about 100k measurements) and some with multiple lattice spacings and volumes to control lattice artifacts. Such programs would have been impossible 5 years ago. As a result, the lattice-QCD tensor charge calculation has the most precise determination of this quantity, which can then be used to constrain the transversity distribution and make predictions for upcoming experiments~\cite{Lin:2017stx}. 
The hyperon couplings are also not precisely known from experiments, and we hope a better determination of these couplings will lead to advancements in multiple subfields.

In this work, we will be using the following definitions for the axial couplings: 
\begin{align}
g_A &= Z_A \langle N |A_{\mu} | N \rangle^{\text{lat}}, \nonumber\\
g_{\Sigma \Sigma} &= Z_A \langle \Sigma |A_{\mu} | \Sigma \rangle^{\text{lat}}/2, \nonumber\\
g_{\Xi \Xi} &= Z_A \langle \Xi |A_{\mu} | \Xi \rangle^{\text{lat}},
\end{align}
where $Z_A$ is the renormalization constant for the axial current. The factor of 2 in $g_{\Sigma \Sigma}$ comes from a Clebsch-Gordan coefficient so that $g_{\Sigma \Sigma} = F$ in the SU(3) limit. 
The hyperon axial structure can be obtained through the following
\begin{multline}
\langle B | A_{\mu}(q) | B \rangle = \\
  u_B(p') \left[ \gamma_\mu \gamma_5 G_A(q^2) + \gamma_5 q_\mu \frac{G_P(q^2)}{2M_B} \right]  u_B(p) e^{-iqx}, 
\end{multline}
where $B$ is an octet baryon ($N$, $\Xi$, $\Sigma$), $u_B$ is the Dirac spinor, $G_A$ is the axial form factor, $G_P$ is the induced pseudoscalar form factor, and $q=p'-p$ is the transfer momentum.
In the $q^2=0$ limit, we obtain the hyperon coupling constants that come from $G_A(q^2=0)$.

There have been several previous LQCD calculations of the hyperon coupling constants.
The first such calculation was performed in 2007 with a single lattice spacing and lightest pion mass near 350~MeV~\cite{Lin:2007ap} using 2+1-flavor lattices; they got $g_{\Sigma \Sigma}=0.450(21)_\text{stat}(27)_\text{sys}$ and $g_{\Xi \Xi}=-0.277(15)_\text{stat}(19)_\text{sys}$.
A follow-up study by a Japanese group used 2-flavor lattices~\cite{Erkol:2009ev}, producing results consistent with the heavier pion masses of Ref.~\cite{Lin:2007ap}.  
ETMC~\cite{Alexandrou:2016xok} used $N_f=2+1+1$ lattices with lowest pion mass 213~MeV and 2 lattice spacings, 0.082 and 0.065~fm; they obtained 
 $g_{\Sigma \Sigma} = 0.381(11)$ and $g_{\Xi \Xi} = -0.248(9)$ (statistical errors only) after extrapolating to the physical pion mass.
In this work, we not only present the first calculation of these quantities at physical pion mass, but also study the finite-volume effects and lattice discretization systematics, and report the first continuum-limit results for the hyperon axial couplings.

\section{Lattice-QCD Calculation Setup}

In this work, we use clover lattice fermion action for the valence quarks on top of 2+1+1 flavors of hypercubic (HYP)-smeared~\cite{Hasenfratz:2001hp} 
highly improved staggered quarks (HISQ)~\cite{Follana:2006rc,Bazavov:2012xda}
in configurations generated by MILC Collaboration. The quark mass and clover parameters are the same as those used by PNDME Collaboration~\cite{Gupta:2018qil}.
We use 3 lattice spacings $a$: 0.06, 0.09 and 0.12~fm with pion masses $M_\pi$ ranging from near physical pion mass (135~MeV) to heaviest ones around 310~MeV. We also perform a volume-dependence study at $a\approx 0.12$~fm and $M_\pi \approx 220$~MeV where $M_\pi L$ ranges from 3.3 to 5.5. A summary of the ensemble information and parameters used in our calculations can be found in Table~\ref{tab:params}.

\begin{table}[tbp]
\begin{center}
\begin{tabular}{|l|ccccc|}
\hline
Ensemble ID & $L_s^3 \times L_t$& $M_\pi^\text{val} L_s$ & $t_\text{sep}/a$  &  $N_\text{conf}$ &  $N_\text{meas}$  \\
\hline
a12m310 & $24^3\times 64$ & 4.5 & $\{8, 9, 10, 11, 12\}$  &    1013 & 4052 \\
a12m220S & $24^3\times 64$ & 3.3 & $\{8, 10,12\}$      &  1000 & 6000  \\
a12m220 & $32^3\times 64$ & 4.4 & $\{8, 10,12\}$   &  958 & 3832 \\
a12m220L & $40^3\times 64$ & 5.5 & $\{10\}$      & 1010 & 4040  \\
a09m310 & $32^3\times 64$ & 4.5 & $\{10, 12, 14\}$  & 775    &  3100  \\
a09m220 & $48^3\times 64$ & 4.8 & $\{10, 12, 14\}$  & 890    &  3560  \\
a09m130 & $64^3\times 64$ & 3.9 & $\{10, 12\}$      & 1058 &  4232  \\
a06m310 & $48^3\times 64$ & 4.5 & $\{13, 16, 20\}$  & 480   & 1920   \\
\hline
\end{tabular}
\end{center}
\caption{\label{tab:params}
Ensemble information and parameters used in this calculation.}
\end{table}

To extract the octet axial couplings, we simultaneously fit the octet two-point ($C_\text{2pt}$) and three point ($C_\text{3pt}$) correlators, including the first
excited state, with the following fit forms:
\begin{equation}
C^\text{2pt} (t) ={|{\cal A}_0|}^2 e^{-M_0 (t)} + {|{\cal A}_1|}^2 e^{-M_1 (t)}\,,
\end{equation}
\begin{align}
\label{eq:c3ptfitform}
C^\text{3pt}_{\Gamma}(t,t_\text{sep}) &= 
   |{\cal A}_0|^2 \langle 0 | \mathcal{O}_\Gamma | 0 \rangle  e^{-M_0t_\text{sep}} \nonumber\\
   &+|{\cal A}_1|^2 \langle 1 | \mathcal{O}_\Gamma | 1 \rangle  e^{-M_1t_\text{sep}} \nonumber\\
   &+{\cal A}_1{\cal A}_0^* \langle 1 | \mathcal{O}_\Gamma | 0 \rangle  e^{-M_1 (t_\text{sep}-t)} e^{-M_0 t} \nonumber\\
   &+{\cal A}_0{\cal A}_1^* \langle 0 | \mathcal{O}_\Gamma | 1 \rangle  e^{-M_0 (t_\text{sep}-t)} e^{-M_1 t} \nonumber\\
&+ \ldots \,
\end{align}
$A_0$ and $A_1$ are overlap amplitudes for the ground and excited states, 
$t_\text{sep}$ is the source-sink separation, 
and $M_0$ and $M_1$ are the masses for ground and excited states of the corresponding octet baryons. For $q^2=0$, the third and fourth terms are related, so Eq.~\ref{eq:c3ptfitform} can be combined into a three-term fit using all the $t_\text{sep}$ data we have. Note that the unwanted matrix element $\langle 1 | \mathcal{O}_\Gamma | 1 \rangle $ has the same time dependence as the wanted ground-state matrix element $\langle 0 | \mathcal{O}_\Gamma | 0 \rangle$, so this excited-state contamination can only be reliably extracted when there are multiple $t_\text{sep}$ in the data. 
We fit the matrix elements from all ensembles up to $\langle 1 | \mathcal{O}_\Gamma | 1 \rangle $ with the exception of the a12m220L ensemble for which only one source-sink separation was taken; only two-term fits (up to $\langle 1 | \mathcal{O}_\Gamma | 0 \rangle$) can be used to extract the bare matrix elements. Based on the measurements using the same lattice spacing and pion mass from the smaller-volume a12m220 ensembles, there is no sign of the excited-state contamination at the chosen source-sink separation.

\begin{figure}[htbp]
\includegraphics[width=.45\textwidth]{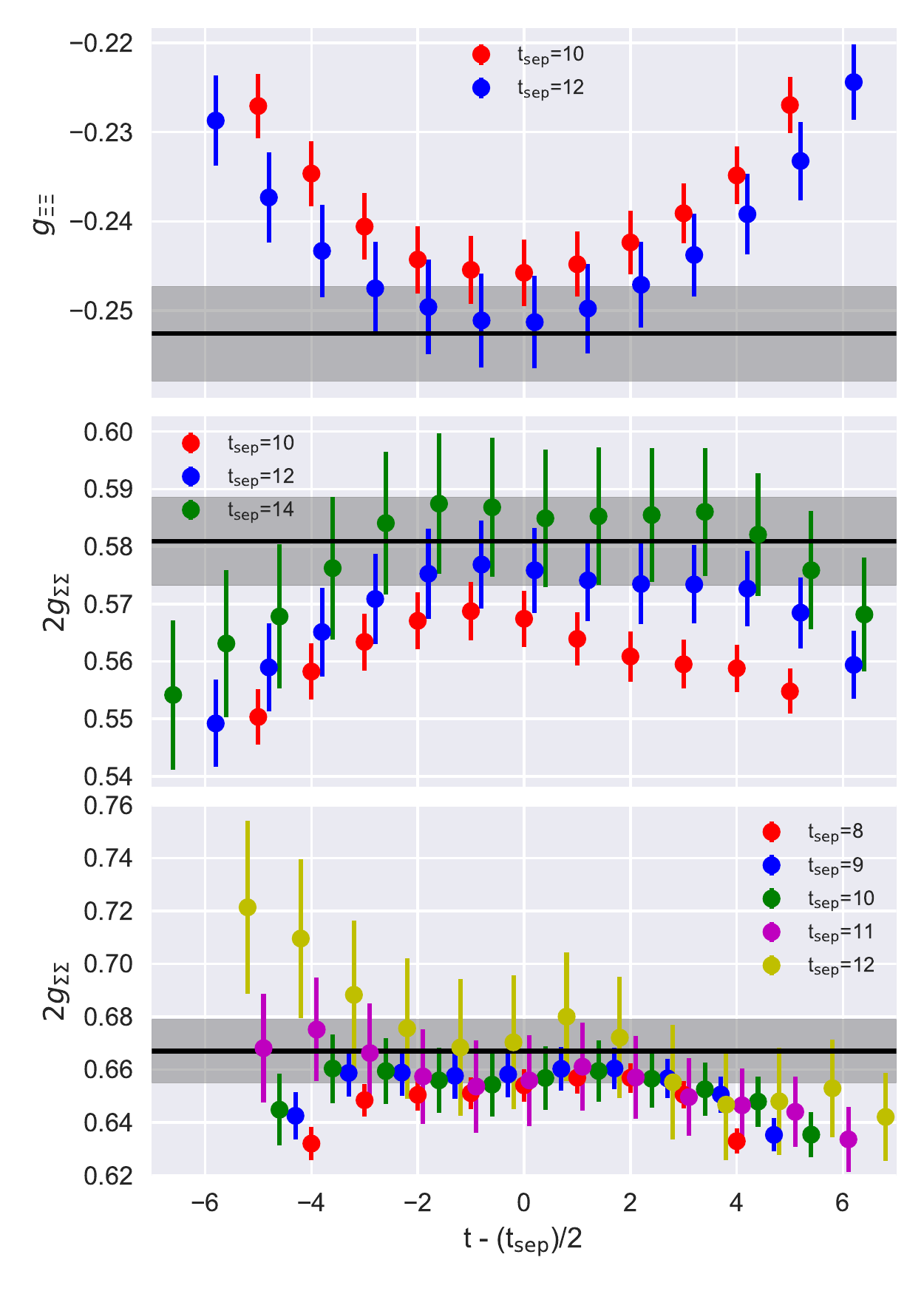}
\caption{Example ratio plots of $C_\text{3pt}/C_\text{2pt}(t_\text{sep})$ for $g_{\Xi \Xi}$ (top, for a09m130 ensemble) and $g_{\Sigma \Sigma}$ (middle and bottom figures, for a09m220 and a12m310) from selected ensembles. The extracted ground-state hyperon matrix elements from two-state fit using all the $t_\text{sep}$ are marked by the black line and gray band.}
\end{figure}

\section{Results and Discussion}

\begin{figure}[htbp]
\includegraphics[width=.45\textwidth]{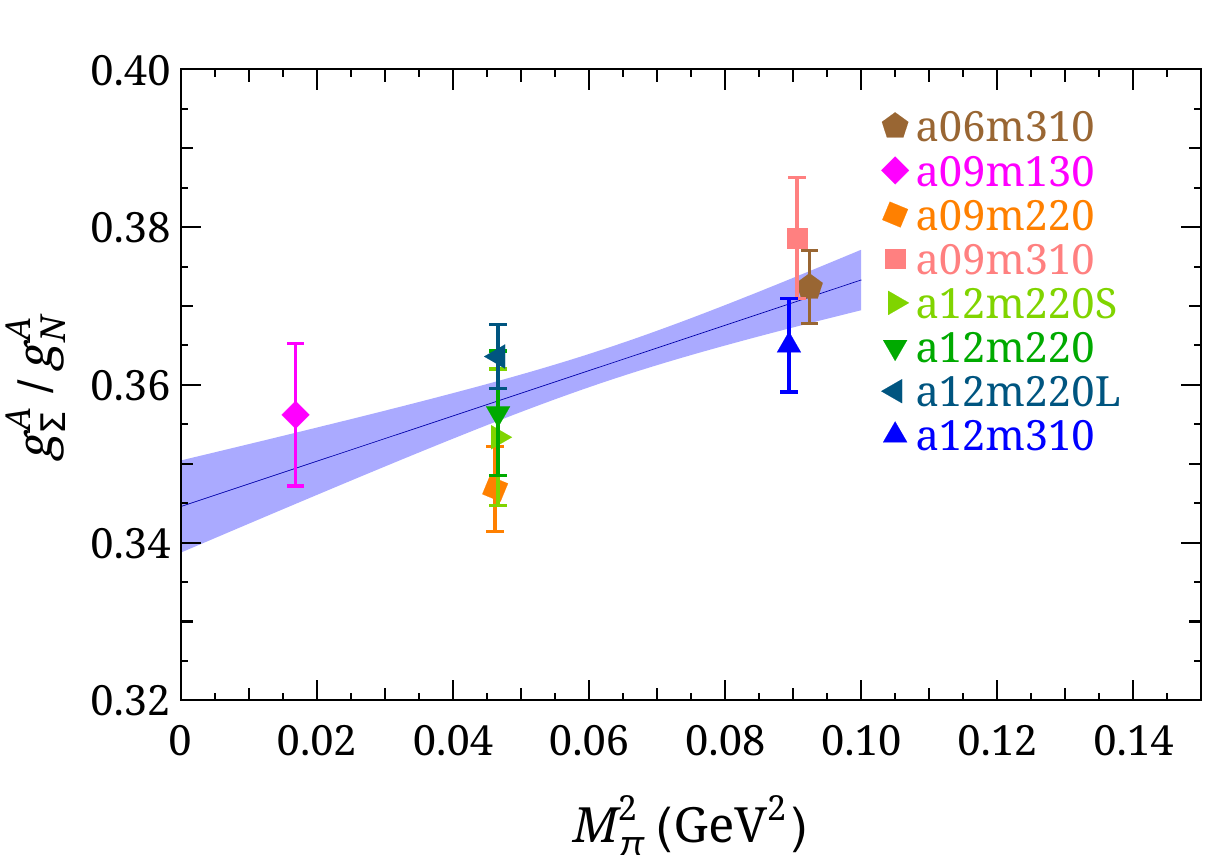}
\includegraphics[width=.45\textwidth]{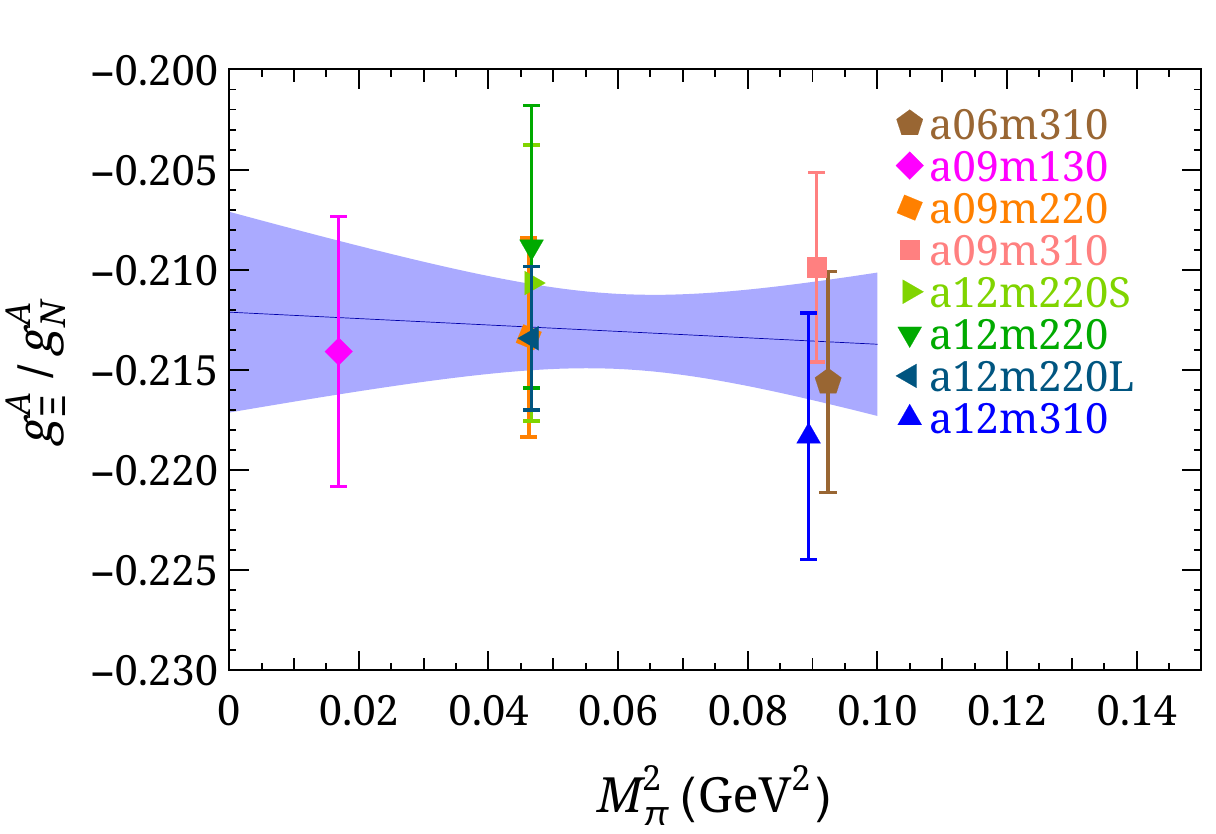}
\caption{Ratios of the $\Sigma$ and $\Xi$ axial coupling to the nucleon axial coupling as functions of pion mass squared. The blue band here shows our favored continuum extrapolation with linear dependence on $M_\pi^2$. 
} \label{fig:sigmaxiextrap}
\end{figure}

Figure~\ref{fig:sigmaxiextrap} summarizes our results as functions of $M_\pi^2$. We choose to use ratios of hyperon to nucleon axial couplings for the following reasons: 
First, we avoid the need to determine the renormalization constants of the axial current, so there is no need to fold in the additional uncertainty due to nonperturbative renormalization. 
Secondly, the signal-to-noise of the ratios are significantly improved due to the correlations in the data, since they are taken using the same QCD configurations. 
Thirdly, we expect some of the lattice artifacts to be canceled or reduced in the ratio combinations. 
As shown in Fig.~\ref{fig:sigmaxiextrap}, we do not see strong pion mass, nor significant lattice-spacing or volume dependence in these ratios. 

A naive continuum extrapolation, the strategy taken by many lattice works in the past, is to assume extrapolation linear in $M_\pi^2$ and neglect the lattice spacing $a$ and volume dependence (using the dimensionless parameter ($M_\pi L$). 
The blue band in Fig.~\ref{fig:sigmaxiextrap} shows such an extrapolation; we obtain $g_{\Sigma \Sigma}/g_A=0.3501(43)$ and $g_{\Xi \Xi}/g_A=-0.2124(37)$. 
Next, we explore the systematics using polynomials in $M_\pi^2$ up to $M_\pi^4$. For each for, we also consider volume dependence of $\{1, e^{-M_\pi L}\}$ and lattice-spacing dependence $\{1, a, a^2\}$. This yields 18 possible continuum-extrapolation forms.
The results of these fits are then combined using the Akaike information criterion (AIC) weighted by the factor
$w_i = P_i/(\sum_j P_j)$,
where $P_j = \exp[-(\text{AIC}_j - \min\text{AIC})/2]$, 
$\text{AIC} = 2k + \chi^2$, 
$k$ is the number of free parameters in a fit,
and $\chi^2$ is the minimum sum of squared fit residuals. 
We then estimate the contribution of systematic uncertainty by taking the difference total error from the statistical-only result (difference  in quadrature); this gives
$g_{\Sigma \Sigma}/g_A=0.3501(43)_\text{stat}(51)_\text{sys}$ and
$g_{\Xi \Xi}/g_A     =-0.2124(37)_\text{stat}(10)_\text{sys}$.
 
Taking the experimental averaged $g_A$ from the Particle Data Guide~\cite{Tanabashi:2018oca}, our final hyperon axial couplings are
$g_{\Sigma \Sigma}=0.4455(55)_\text{stat}(65)_\text{sys}$ and
$g_{\Xi \Xi}     =-0.2703(47)_\text{stat}(13)_\text{sys}$.
Our results are consistent with the first hyperon axial coupling calculation~\cite{Lin:2007ap}, 
$g_{\Sigma \Sigma}=0.450(21)_\text{stat}(27)_\text{sys}$ and
$g_{\Xi \Xi}=-0.277(15)_\text{stat}(19)_\text{sys}$
(based on a single lattice spacing and much heavier pion masses), but we achieve much improvement in statistical uncertainty and control of systematics. 
Assuming SU(3) symmetry where $g_{\Sigma \Sigma}=F$ and $g_{\Xi \Xi}=D-F$, we obtained low-energy chiral parameters
$D = 0.708(10)_\text{stat}(6)_\text{sys}$ and
$F = 0.438(7)_\text{stat}(6)_\text{sys}$,
which are not consistent with the determination $D = 0.804(8)$ and
$F = 0.463(8)$ using semileptonic decay data under the same assumption of SU(3) symmetry~\cite{Cabibbo:2003cu}.
Using low-energy constants, we can determine 
$a_8= 3F- D = 0.607(16)_\text{stat}(12)_\text{sys}$,
which gives the proton spin structure of $\Delta u + \Delta d -2 \Delta s$,
consistent with the value of 0.585(25) commonly used in polarized global fits as a constraint~\cite{Lin:2017snn}.

\begin{figure}[htbp]
\includegraphics[width=.45\textwidth]{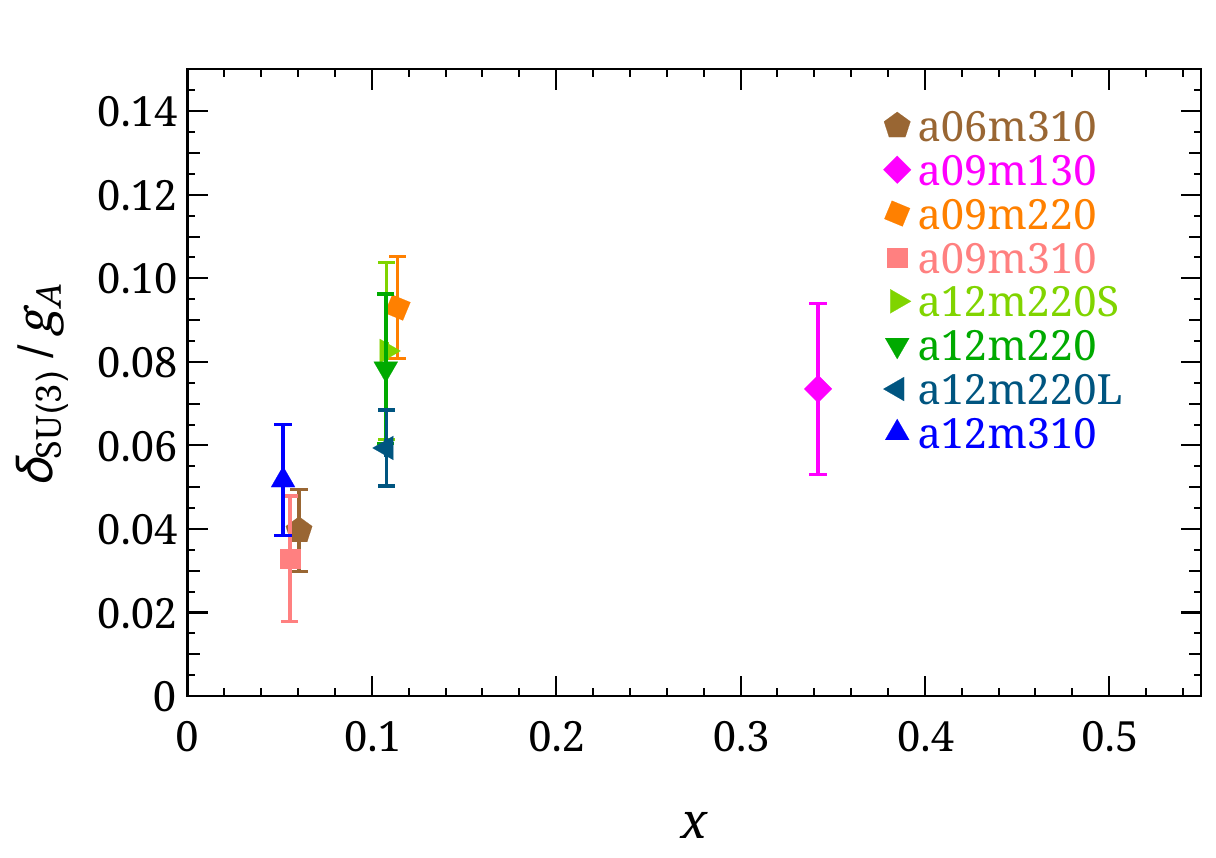}
\caption{
The ratios of $SU(3)$ symmetry breaking ($\delta_\text{SU(3)}$) to nucleon axial charge ($g_A$) as a function of $x=(M_K^2-M_\pi^2)/(4\pi M_\pi^2)$. 
There is a noticeable increase with $x$ at heavier quark mass but it is mild when only looking at the lightest 2 pion masses. 
} \label{fig:su3}
\end{figure}

We also investigate study the extent of SU(3) symmetry breaking by considering the quantity 
\begin{equation} 
\delta_\text{SU(3)} = g_A -2 g_{\Sigma \Sigma}+g_{\Xi \Xi}.
\end{equation} 
Figure~\ref{fig:su3} shows $\delta_\text{SU(3)}/g_A$ for all the ensembles used in this work as a function of the dimensionless quantity
$x = (M_K^2-M_\pi^2)/(4\pi M_\pi^2)$ measured independently on each ensemble.  
In contrast to the earlier works~\cite{Lin:2007ap,Erkol:2009ev,Alexandrou:2016xok}, we do not see a strong dependence of $\delta_\text{SU(3)}$ proportional to the $x^2$ for pion masses between 220~MeV and the physical pion mass.
This is likely due to the heavier pion masses used in the previous studies~\cite{Lin:2007ap,Erkol:2009ev,Alexandrou:2016xok}; the dependence is more obvious when we only look at the heaviest 2 pion masses, but it disappears for pion mass below 220~MeV. 

Using the $g_{\Sigma \Sigma}$ and $g_{\Xi \Xi}$ found earlier, we obtain $\delta_\text{SU(3)}/g_A=+0.087(10)_\text{stat}(11)_\text{sys}$.
Thus, we estimate a total SU(3) symmetry breaking size of about 9\%, which is smaller than the estimate from Ref.~\cite{Lin:2007ap}. 

\section{Summary and Outlook}

In this work, we have calculated the axial coupling of the $\Sigma$ and $\Xi$ octet baryons using $N_f=2+1+1$ lattice QCD. For the first time, not only have these quantities been studied directly at the physical pion mass but also with careful study of the sources of systematic uncertainty, including the lattice-spacing and the finite-volume effects. We calculated multiple source-sink separations for the three-point correlators and used
a two-state strategy to fit all separation data simultaneously to remove excited-state contamination. 
We constructed the ratios of $g_{\Sigma \Sigma}/g_A$ and $g_{\Xi \Xi}/g_A$ and found these ratios have smaller dependence on the $M_\pi^2$, lattice spacing $a$ and volume. 
We then extrapolated the ratios to the physical limit using 18 different fitting forms and obtained 
$g_{\Sigma \Sigma}=0.4455(55)_\text{stat}(65)_\text{sys}$ and
$g_{\Xi \Xi}     =-0.2703(47)_\text{stat}(13)_\text{sys}$.
We also examined the SU(3) symmetry breaking using these couplings and found around 9\% effect in this updated study. 

\section*{Acknowledgments}

We thank the MILC Collaboration for sharing the lattices used to perform this study; the LQCD calculations were performed using the Chroma software suite~\cite{Edwards:2004sx}. This work is supported by Michigan State University through computational resources provided by the Institute for Cyber-Enabled Research. HL is supported by the US National Science Foundation under grant PHY 1653405 ``CAREER: Constraining Parton Distribution Functions for New-Physics Searches''.


\begin{thebibliography}{25}%
\makeatletter
\providecommand \@ifxundefined [1]{%
 \@ifx{#1\undefined}
}%
\providecommand \@ifnum [1]{%
 \ifnum #1\expandafter \@firstoftwo
 \else \expandafter \@secondoftwo
 \fi
}%
\providecommand \@ifx [1]{%
 \ifx #1\expandafter \@firstoftwo
 \else \expandafter \@secondoftwo
 \fi
}%
\providecommand \natexlab [1]{#1}%
\providecommand \enquote  [1]{``#1''}%
\providecommand \bibnamefont  [1]{#1}%
\providecommand \bibfnamefont [1]{#1}%
\providecommand \citenamefont [1]{#1}%
\providecommand \href@noop [0]{\@secondoftwo}%
\providecommand \href [0]{\begingroup \@sanitize@url \@href}%
\providecommand \@href[1]{\@@startlink{#1}\@@href}%
\providecommand \@@href[1]{\endgroup#1\@@endlink}%
\providecommand \@sanitize@url [0]{\catcode `\\12\catcode `\$12\catcode
  `\&12\catcode `\#12\catcode `\^12\catcode `\_12\catcode `\%12\relax}%
\providecommand \@@startlink[1]{}%
\providecommand \@@endlink[0]{}%
\providecommand \url  [0]{\begingroup\@sanitize@url \@url }%
\providecommand \@url [1]{\endgroup\@href {#1}{\urlprefix }}%
\providecommand \urlprefix  [0]{URL }%
\providecommand \Eprint [0]{\href }%
\providecommand \doibase [0]{http://dx.doi.org/}%
\providecommand \selectlanguage [0]{\@gobble}%
\providecommand \bibinfo  [0]{\@secondoftwo}%
\providecommand \bibfield  [0]{\@secondoftwo}%
\providecommand \translation [1]{[#1]}%
\providecommand \BibitemOpen [0]{}%
\providecommand \bibitemStop [0]{}%
\providecommand \bibitemNoStop [0]{.\EOS\space}%
\providecommand \EOS [0]{\spacefactor3000\relax}%
\providecommand \BibitemShut  [1]{\csname bibitem#1\endcsname}%
\let\auto@bib@innerbib\@empty
\bibitem [{\citenamefont {Savage}\ and\ \citenamefont
  {Walden}(1997)}]{Savage:1996zd}%
  \BibitemOpen
  \bibfield  {author} {\bibinfo {author} {\bibfnamefont {M.~J.}\ \bibnamefont
  {Savage}}\ and\ \bibinfo {author} {\bibfnamefont {J.}~\bibnamefont
  {Walden}},\ }\href {\doibase 10.1103/PhysRevD.55.5376} {\bibfield  {journal}
  {\bibinfo  {journal} {Phys. Rev.}\ }\textbf {\bibinfo {volume} {D55}},\
  \bibinfo {pages} {5376} (\bibinfo {year} {1997})},\ \Eprint
  {http://arxiv.org/abs/hep-ph/9611210} {arXiv:hep-ph/9611210 [hep-ph]}
  \BibitemShut {NoStop}%
\bibitem [{\citenamefont {Cabibbo}\ \emph {et~al.}(2003)\citenamefont
  {Cabibbo}, \citenamefont {Swallow},\ and\ \citenamefont
  {Winston}}]{Cabibbo:2003cu}%
  \BibitemOpen
  \bibfield  {author} {\bibinfo {author} {\bibfnamefont {N.}~\bibnamefont
  {Cabibbo}}, \bibinfo {author} {\bibfnamefont {E.~C.}\ \bibnamefont
  {Swallow}}, \ and\ \bibinfo {author} {\bibfnamefont {R.}~\bibnamefont
  {Winston}},\ }\href {\doibase 10.1146/annurev.nucl.53.013103.155258}
  {\bibfield  {journal} {\bibinfo  {journal} {Ann. Rev. Nucl. Part. Sci.}\
  }\textbf {\bibinfo {volume} {53}},\ \bibinfo {pages} {39} (\bibinfo {year}
  {2003})},\ \Eprint {http://arxiv.org/abs/hep-ph/0307298}
  {arXiv:hep-ph/0307298 [hep-ph]} \BibitemShut {NoStop}%
\bibitem [{\citenamefont {Beane}\ \emph {et~al.}(2005)\citenamefont {Beane},
  \citenamefont {Bedaque}, \citenamefont {Parreno},\ and\ \citenamefont
  {Savage}}]{Beane:2003yx}%
  \BibitemOpen
  \bibfield  {author} {\bibinfo {author} {\bibfnamefont {S.~R.}\ \bibnamefont
  {Beane}}, \bibinfo {author} {\bibfnamefont {P.~F.}\ \bibnamefont {Bedaque}},
  \bibinfo {author} {\bibfnamefont {A.}~\bibnamefont {Parreno}}, \ and\
  \bibinfo {author} {\bibfnamefont {M.~J.}\ \bibnamefont {Savage}},\ }\href
  {\doibase 10.1016/j.nuclphysa.2004.09.081} {\bibfield  {journal} {\bibinfo
  {journal} {Nucl. Phys.}\ }\textbf {\bibinfo {volume} {A747}},\ \bibinfo
  {pages} {55} (\bibinfo {year} {2005})},\ \Eprint
  {http://arxiv.org/abs/nucl-th/0311027} {arXiv:nucl-th/0311027 [nucl-th]}
  \BibitemShut {NoStop}%
\bibitem [{\citenamefont {Lattimer}\ and\ \citenamefont
  {Prakash}(2007)}]{Lattimer:2006xb}%
  \BibitemOpen
  \bibfield  {author} {\bibinfo {author} {\bibfnamefont {J.~M.}\ \bibnamefont
  {Lattimer}}\ and\ \bibinfo {author} {\bibfnamefont {M.}~\bibnamefont
  {Prakash}},\ }\href {\doibase 10.1016/j.physrep.2007.02.003} {\bibfield
  {journal} {\bibinfo  {journal} {Phys. Rept.}\ }\textbf {\bibinfo {volume}
  {442}},\ \bibinfo {pages} {109} (\bibinfo {year} {2007})},\ \Eprint
  {http://arxiv.org/abs/astro-ph/0612440} {arXiv:astro-ph/0612440 [astro-ph]}
  \BibitemShut {NoStop}%
\bibitem [{\citenamefont {Weissenborn}\ \emph {et~al.}(2012)\citenamefont
  {Weissenborn}, \citenamefont {Chatterjee},\ and\ \citenamefont
  {Schaffner-Bielich}}]{Weissenborn:2011ut}%
  \BibitemOpen
  \bibfield  {author} {\bibinfo {author} {\bibfnamefont {S.}~\bibnamefont
  {Weissenborn}}, \bibinfo {author} {\bibfnamefont {D.}~\bibnamefont
  {Chatterjee}}, \ and\ \bibinfo {author} {\bibfnamefont {J.}~\bibnamefont
  {Schaffner-Bielich}},\ }\href {\doibase 10.1103/PhysRevC.85.065802,
  10.1103/PhysRevC.90.019904} {\bibfield  {journal} {\bibinfo  {journal} {Phys.
  Rev.}\ }\textbf {\bibinfo {volume} {C85}},\ \bibinfo {pages} {065802}
  (\bibinfo {year} {2012})},\ \bibinfo {note} {[Erratum: Phys.
  Rev.C90,no.1,019904(2014)]},\ \Eprint {http://arxiv.org/abs/1112.0234}
  {arXiv:1112.0234 [astro-ph.HE]} \BibitemShut {NoStop}%
\bibitem [{\citenamefont {Dai}\ \emph {et~al.}(1996)\citenamefont {Dai},
  \citenamefont {Dashen}, \citenamefont {Jenkins},\ and\ \citenamefont
  {Manohar}}]{Dai:1995zg}%
  \BibitemOpen
  \bibfield  {author} {\bibinfo {author} {\bibfnamefont {J.}~\bibnamefont
  {Dai}}, \bibinfo {author} {\bibfnamefont {R.~F.}\ \bibnamefont {Dashen}},
  \bibinfo {author} {\bibfnamefont {E.~E.}\ \bibnamefont {Jenkins}}, \ and\
  \bibinfo {author} {\bibfnamefont {A.~V.}\ \bibnamefont {Manohar}},\ }\href
  {\doibase 10.1103/PhysRevD.53.273} {\bibfield  {journal} {\bibinfo  {journal}
  {Phys. Rev.}\ }\textbf {\bibinfo {volume} {D53}},\ \bibinfo {pages} {273}
  (\bibinfo {year} {1996})},\ \Eprint {http://arxiv.org/abs/hep-ph/9506273}
  {arXiv:hep-ph/9506273 [hep-ph]} \BibitemShut {NoStop}%
\bibitem [{\citenamefont {Lin}\ \emph {et~al.}(2018{\natexlab{a}})\citenamefont
  {Lin} \emph {et~al.}}]{Lin:2017snn}%
  \BibitemOpen
  \bibfield  {author} {\bibinfo {author} {\bibfnamefont {H.-W.}\ \bibnamefont
  {Lin}} \emph {et~al.},\ }\href {\doibase 10.1016/j.ppnp.2018.01.007}
  {\bibfield  {journal} {\bibinfo  {journal} {Prog. Part. Nucl. Phys.}\
  }\textbf {\bibinfo {volume} {100}},\ \bibinfo {pages} {107} (\bibinfo {year}
  {2018}{\natexlab{a}})},\ \Eprint {http://arxiv.org/abs/1711.07916}
  {arXiv:1711.07916 [hep-ph]} \BibitemShut {NoStop}%
\bibitem [{\citenamefont {Gupta}\ \emph {et~al.}(2018)\citenamefont {Gupta},
  \citenamefont {Jang}, \citenamefont {Yoon}, \citenamefont {Lin},
  \citenamefont {Cirigliano},\ and\ \citenamefont
  {Bhattacharya}}]{Gupta:2018qil}%
  \BibitemOpen
  \bibfield  {author} {\bibinfo {author} {\bibfnamefont {R.}~\bibnamefont
  {Gupta}}, \bibinfo {author} {\bibfnamefont {Y.-C.}\ \bibnamefont {Jang}},
  \bibinfo {author} {\bibfnamefont {B.}~\bibnamefont {Yoon}}, \bibinfo {author}
  {\bibfnamefont {H.-W.}\ \bibnamefont {Lin}}, \bibinfo {author} {\bibfnamefont
  {V.}~\bibnamefont {Cirigliano}}, \ and\ \bibinfo {author} {\bibfnamefont
  {T.}~\bibnamefont {Bhattacharya}},\ }\href {\doibase
  10.1103/PhysRevD.98.034503} {\bibfield  {journal} {\bibinfo  {journal} {Phys.
  Rev.}\ }\textbf {\bibinfo {volume} {D98}},\ \bibinfo {pages} {034503}
  (\bibinfo {year} {2018})},\ \Eprint {http://arxiv.org/abs/1806.09006}
  {arXiv:1806.09006 [hep-lat]} \BibitemShut {NoStop}%
\bibitem [{\citenamefont {Lin}\ \emph {et~al.}(2018{\natexlab{b}})\citenamefont
  {Lin}, \citenamefont {Chen}, \citenamefont {Jin}, \citenamefont {Liu},
  \citenamefont {Yang}, \citenamefont {Zhang},\ and\ \citenamefont
  {Zhao}}]{Lin:2018qky}%
  \BibitemOpen
  \bibfield  {author} {\bibinfo {author} {\bibfnamefont {H.-W.}\ \bibnamefont
  {Lin}}, \bibinfo {author} {\bibfnamefont {J.-W.}\ \bibnamefont {Chen}},
  \bibinfo {author} {\bibfnamefont {L.}~\bibnamefont {Jin}}, \bibinfo {author}
  {\bibfnamefont {Y.-S.}\ \bibnamefont {Liu}}, \bibinfo {author} {\bibfnamefont
  {Y.-B.}\ \bibnamefont {Yang}}, \bibinfo {author} {\bibfnamefont {J.-H.}\
  \bibnamefont {Zhang}}, \ and\ \bibinfo {author} {\bibfnamefont
  {Y.}~\bibnamefont {Zhao}},\ }\href@noop {} {\  (\bibinfo {year}
  {2018}{\natexlab{b}})},\ \Eprint {http://arxiv.org/abs/1807.07431}
  {arXiv:1807.07431 [hep-lat]} \BibitemShut {NoStop}%
\bibitem [{\citenamefont {Bhattacharya}\ \emph {et~al.}(2016)\citenamefont
  {Bhattacharya}, \citenamefont {Cirigliano}, \citenamefont {Cohen},
  \citenamefont {Gupta}, \citenamefont {Lin},\ and\ \citenamefont
  {Yoon}}]{Bhattacharya:2016zcn}%
  \BibitemOpen
  \bibfield  {author} {\bibinfo {author} {\bibfnamefont {T.}~\bibnamefont
  {Bhattacharya}}, \bibinfo {author} {\bibfnamefont {V.}~\bibnamefont
  {Cirigliano}}, \bibinfo {author} {\bibfnamefont {S.}~\bibnamefont {Cohen}},
  \bibinfo {author} {\bibfnamefont {R.}~\bibnamefont {Gupta}}, \bibinfo
  {author} {\bibfnamefont {H.-W.}\ \bibnamefont {Lin}}, \ and\ \bibinfo
  {author} {\bibfnamefont {B.}~\bibnamefont {Yoon}},\ }\href {\doibase
  10.1103/PhysRevD.94.054508} {\bibfield  {journal} {\bibinfo  {journal} {Phys.
  Rev.}\ }\textbf {\bibinfo {volume} {D94}},\ \bibinfo {pages} {054508}
  (\bibinfo {year} {2016})},\ \Eprint {http://arxiv.org/abs/1606.07049}
  {arXiv:1606.07049 [hep-lat]} \BibitemShut {NoStop}%
\bibitem [{\citenamefont {Bhattacharya}\ \emph {et~al.}(2015)\citenamefont
  {Bhattacharya}, \citenamefont {Cirigliano}, \citenamefont {Cohen},
  \citenamefont {Gupta}, \citenamefont {Joseph}, \citenamefont {Lin},\ and\
  \citenamefont {Yoon}}]{Bhattacharya:2015wna}%
  \BibitemOpen
  \bibfield  {author} {\bibinfo {author} {\bibfnamefont {T.}~\bibnamefont
  {Bhattacharya}}, \bibinfo {author} {\bibfnamefont {V.}~\bibnamefont
  {Cirigliano}}, \bibinfo {author} {\bibfnamefont {S.}~\bibnamefont {Cohen}},
  \bibinfo {author} {\bibfnamefont {R.}~\bibnamefont {Gupta}}, \bibinfo
  {author} {\bibfnamefont {A.}~\bibnamefont {Joseph}}, \bibinfo {author}
  {\bibfnamefont {H.-W.}\ \bibnamefont {Lin}}, \ and\ \bibinfo {author}
  {\bibfnamefont {B.}~\bibnamefont {Yoon}} (\bibinfo {collaboration} {PNDME}),\
  }\href {\doibase 10.1103/PhysRevD.92.094511} {\bibfield  {journal} {\bibinfo
  {journal} {Phys. Rev.}\ }\textbf {\bibinfo {volume} {D92}},\ \bibinfo {pages}
  {094511} (\bibinfo {year} {2015})},\ \Eprint
  {http://arxiv.org/abs/1506.06411} {arXiv:1506.06411 [hep-lat]} \BibitemShut
  {NoStop}%
\bibitem [{\citenamefont {Green}\ \emph {et~al.}(2012)\citenamefont {Green},
  \citenamefont {Negele}, \citenamefont {Pochinsky}, \citenamefont {Syritsyn},
  \citenamefont {Engelhardt},\ and\ \citenamefont {Krieg}}]{Green:2012ej}%
  \BibitemOpen
  \bibfield  {author} {\bibinfo {author} {\bibfnamefont {J.~R.}\ \bibnamefont
  {Green}}, \bibinfo {author} {\bibfnamefont {J.~W.}\ \bibnamefont {Negele}},
  \bibinfo {author} {\bibfnamefont {A.~V.}\ \bibnamefont {Pochinsky}}, \bibinfo
  {author} {\bibfnamefont {S.~N.}\ \bibnamefont {Syritsyn}}, \bibinfo {author}
  {\bibfnamefont {M.}~\bibnamefont {Engelhardt}}, \ and\ \bibinfo {author}
  {\bibfnamefont {S.}~\bibnamefont {Krieg}},\ }\href {\doibase
  10.1103/PhysRevD.86.114509} {\bibfield  {journal} {\bibinfo  {journal} {Phys.
  Rev.}\ }\textbf {\bibinfo {volume} {D86}},\ \bibinfo {pages} {114509}
  (\bibinfo {year} {2012})},\ \Eprint {http://arxiv.org/abs/1206.4527}
  {arXiv:1206.4527 [hep-lat]} \BibitemShut {NoStop}%
\bibitem [{\citenamefont {Aoki}\ \emph {et~al.}(2010)\citenamefont {Aoki},
  \citenamefont {Blum}, \citenamefont {Lin}, \citenamefont {Ohta},
  \citenamefont {Sasaki}, \citenamefont {Tweedie}, \citenamefont {Zanotti},\
  and\ \citenamefont {Yamazaki}}]{Aoki:2010xg}%
  \BibitemOpen
  \bibfield  {author} {\bibinfo {author} {\bibfnamefont {Y.}~\bibnamefont
  {Aoki}}, \bibinfo {author} {\bibfnamefont {T.}~\bibnamefont {Blum}}, \bibinfo
  {author} {\bibfnamefont {H.-W.}\ \bibnamefont {Lin}}, \bibinfo {author}
  {\bibfnamefont {S.}~\bibnamefont {Ohta}}, \bibinfo {author} {\bibfnamefont
  {S.}~\bibnamefont {Sasaki}}, \bibinfo {author} {\bibfnamefont
  {R.}~\bibnamefont {Tweedie}}, \bibinfo {author} {\bibfnamefont
  {J.}~\bibnamefont {Zanotti}}, \ and\ \bibinfo {author} {\bibfnamefont
  {T.}~\bibnamefont {Yamazaki}},\ }\href {\doibase 10.1103/PhysRevD.82.014501}
  {\bibfield  {journal} {\bibinfo  {journal} {Phys. Rev.}\ }\textbf {\bibinfo
  {volume} {D82}},\ \bibinfo {pages} {014501} (\bibinfo {year} {2010})},\
  \Eprint {http://arxiv.org/abs/1003.3387} {arXiv:1003.3387 [hep-lat]}
  \BibitemShut {NoStop}%
\bibitem [{\citenamefont {Abdel-Rehim}\ \emph {et~al.}(2015)\citenamefont
  {Abdel-Rehim} \emph {et~al.}}]{Abdel-Rehim:2015owa}%
  \BibitemOpen
  \bibfield  {author} {\bibinfo {author} {\bibfnamefont {A.}~\bibnamefont
  {Abdel-Rehim}} \emph {et~al.},\ }\href {\doibase 10.1103/PhysRevD.92.114513,
  10.1103/PhysRevD.93.039904} {\bibfield  {journal} {\bibinfo  {journal} {Phys.
  Rev.}\ }\textbf {\bibinfo {volume} {D92}},\ \bibinfo {pages} {114513}
  (\bibinfo {year} {2015})},\ \bibinfo {note} {[Erratum: Phys.
  Rev.D93,no.3,039904(2016)]},\ \Eprint {http://arxiv.org/abs/1507.04936}
  {arXiv:1507.04936 [hep-lat]} \BibitemShut {NoStop}%
\bibitem [{\citenamefont {Bali}\ \emph {et~al.}(2015)\citenamefont {Bali},
  \citenamefont {Collins}, \citenamefont {Glässle}, \citenamefont {Göckeler},
  \citenamefont {Najjar}, \citenamefont {Rödl}, \citenamefont {Schäfer},
  \citenamefont {Schiel}, \citenamefont {Söldner},\ and\ \citenamefont
  {Sternbeck}}]{Bali:2014nma}%
  \BibitemOpen
  \bibfield  {author} {\bibinfo {author} {\bibfnamefont {G.~S.}\ \bibnamefont
  {Bali}}, \bibinfo {author} {\bibfnamefont {S.}~\bibnamefont {Collins}},
  \bibinfo {author} {\bibfnamefont {B.}~\bibnamefont {Glässle}}, \bibinfo
  {author} {\bibfnamefont {M.}~\bibnamefont {Göckeler}}, \bibinfo {author}
  {\bibfnamefont {J.}~\bibnamefont {Najjar}}, \bibinfo {author} {\bibfnamefont
  {R.~H.}\ \bibnamefont {Rödl}}, \bibinfo {author} {\bibfnamefont
  {A.}~\bibnamefont {Schäfer}}, \bibinfo {author} {\bibfnamefont {R.~W.}\
  \bibnamefont {Schiel}}, \bibinfo {author} {\bibfnamefont {W.}~\bibnamefont
  {Söldner}}, \ and\ \bibinfo {author} {\bibfnamefont {A.}~\bibnamefont
  {Sternbeck}},\ }\href {\doibase 10.1103/PhysRevD.91.054501} {\bibfield
  {journal} {\bibinfo  {journal} {Phys. Rev.}\ }\textbf {\bibinfo {volume}
  {D91}},\ \bibinfo {pages} {054501} (\bibinfo {year} {2015})},\ \Eprint
  {http://arxiv.org/abs/1412.7336} {arXiv:1412.7336 [hep-lat]} \BibitemShut
  {NoStop}%
\bibitem [{\citenamefont {Yamazaki}\ \emph {et~al.}(2008)\citenamefont
  {Yamazaki}, \citenamefont {Aoki}, \citenamefont {Blum}, \citenamefont {Lin},
  \citenamefont {Lin}, \citenamefont {Ohta}, \citenamefont {Sasaki},
  \citenamefont {Tweedie},\ and\ \citenamefont {Zanotti}}]{Yamazaki:2008py}%
  \BibitemOpen
  \bibfield  {author} {\bibinfo {author} {\bibfnamefont {T.}~\bibnamefont
  {Yamazaki}}, \bibinfo {author} {\bibfnamefont {Y.}~\bibnamefont {Aoki}},
  \bibinfo {author} {\bibfnamefont {T.}~\bibnamefont {Blum}}, \bibinfo {author}
  {\bibfnamefont {H.~W.}\ \bibnamefont {Lin}}, \bibinfo {author} {\bibfnamefont
  {M.~F.}\ \bibnamefont {Lin}}, \bibinfo {author} {\bibfnamefont
  {S.}~\bibnamefont {Ohta}}, \bibinfo {author} {\bibfnamefont {S.}~\bibnamefont
  {Sasaki}}, \bibinfo {author} {\bibfnamefont {R.~J.}\ \bibnamefont {Tweedie}},
  \ and\ \bibinfo {author} {\bibfnamefont {J.~M.}\ \bibnamefont {Zanotti}}
  (\bibinfo {collaboration} {RBC+UKQCD}),\ }\href {\doibase
  10.1103/PhysRevLett.100.171602} {\bibfield  {journal} {\bibinfo  {journal}
  {Phys. Rev. Lett.}\ }\textbf {\bibinfo {volume} {100}},\ \bibinfo {pages}
  {171602} (\bibinfo {year} {2008})},\ \Eprint {http://arxiv.org/abs/0801.4016}
  {arXiv:0801.4016 [hep-lat]} \BibitemShut {NoStop}%
\bibitem [{\citenamefont {Lin}\ \emph {et~al.}(2018{\natexlab{c}})\citenamefont
  {Lin}, \citenamefont {Melnitchouk}, \citenamefont {Prokudin}, \citenamefont
  {Sato},\ and\ \citenamefont {Shows}}]{Lin:2017stx}%
  \BibitemOpen
  \bibfield  {author} {\bibinfo {author} {\bibfnamefont {H.-W.}\ \bibnamefont
  {Lin}}, \bibinfo {author} {\bibfnamefont {W.}~\bibnamefont {Melnitchouk}},
  \bibinfo {author} {\bibfnamefont {A.}~\bibnamefont {Prokudin}}, \bibinfo
  {author} {\bibfnamefont {N.}~\bibnamefont {Sato}}, \ and\ \bibinfo {author}
  {\bibfnamefont {H.}~\bibnamefont {Shows}},\ }\href {\doibase
  10.1103/PhysRevLett.120.152502} {\bibfield  {journal} {\bibinfo  {journal}
  {Phys. Rev. Lett.}\ }\textbf {\bibinfo {volume} {120}},\ \bibinfo {pages}
  {152502} (\bibinfo {year} {2018}{\natexlab{c}})},\ \Eprint
  {http://arxiv.org/abs/1710.09858} {arXiv:1710.09858 [hep-ph]} \BibitemShut
  {NoStop}%
\bibitem [{\citenamefont {Lin}\ and\ \citenamefont
  {Orginos}(2009)}]{Lin:2007ap}%
  \BibitemOpen
  \bibfield  {author} {\bibinfo {author} {\bibfnamefont {H.-W.}\ \bibnamefont
  {Lin}}\ and\ \bibinfo {author} {\bibfnamefont {K.}~\bibnamefont {Orginos}},\
  }\href {\doibase 10.1103/PhysRevD.79.034507} {\bibfield  {journal} {\bibinfo
  {journal} {Phys. Rev.}\ }\textbf {\bibinfo {volume} {D79}},\ \bibinfo {pages}
  {034507} (\bibinfo {year} {2009})},\ \Eprint {http://arxiv.org/abs/0712.1214}
  {arXiv:0712.1214 [hep-lat]} \BibitemShut {NoStop}%
\bibitem [{\citenamefont {Erkol}\ \emph {et~al.}(2010)\citenamefont {Erkol},
  \citenamefont {Oka},\ and\ \citenamefont {Takahashi}}]{Erkol:2009ev}%
  \BibitemOpen
  \bibfield  {author} {\bibinfo {author} {\bibfnamefont {G.}~\bibnamefont
  {Erkol}}, \bibinfo {author} {\bibfnamefont {M.}~\bibnamefont {Oka}}, \ and\
  \bibinfo {author} {\bibfnamefont {T.~T.}\ \bibnamefont {Takahashi}},\ }\href
  {\doibase 10.1016/j.physletb.2010.02.016} {\bibfield  {journal} {\bibinfo
  {journal} {Phys. Lett.}\ }\textbf {\bibinfo {volume} {B686}},\ \bibinfo
  {pages} {36} (\bibinfo {year} {2010})},\ \Eprint
  {http://arxiv.org/abs/0911.2447} {arXiv:0911.2447 [hep-lat]} \BibitemShut
  {NoStop}%
\bibitem [{\citenamefont {Alexandrou}\ \emph {et~al.}(2016)\citenamefont
  {Alexandrou}, \citenamefont {Hadjiyiannakou},\ and\ \citenamefont
  {Kallidonis}}]{Alexandrou:2016xok}%
  \BibitemOpen
  \bibfield  {author} {\bibinfo {author} {\bibfnamefont {C.}~\bibnamefont
  {Alexandrou}}, \bibinfo {author} {\bibfnamefont {K.}~\bibnamefont
  {Hadjiyiannakou}}, \ and\ \bibinfo {author} {\bibfnamefont {C.}~\bibnamefont
  {Kallidonis}},\ }\href {\doibase 10.1103/PhysRevD.94.034502} {\bibfield
  {journal} {\bibinfo  {journal} {Phys. Rev.}\ }\textbf {\bibinfo {volume}
  {D94}},\ \bibinfo {pages} {034502} (\bibinfo {year} {2016})},\ \Eprint
  {http://arxiv.org/abs/1606.01650} {arXiv:1606.01650 [hep-lat]} \BibitemShut
  {NoStop}%
\bibitem [{\citenamefont {Hasenfratz}\ and\ \citenamefont
  {Knechtli}(2001)}]{Hasenfratz:2001hp}%
  \BibitemOpen
  \bibfield  {author} {\bibinfo {author} {\bibfnamefont {A.}~\bibnamefont
  {Hasenfratz}}\ and\ \bibinfo {author} {\bibfnamefont {F.}~\bibnamefont
  {Knechtli}},\ }\href {\doibase 10.1103/PhysRevD.64.034504} {\bibfield
  {journal} {\bibinfo  {journal} {Phys. Rev.}\ }\textbf {\bibinfo {volume}
  {D64}},\ \bibinfo {pages} {034504} (\bibinfo {year} {2001})},\ \Eprint
  {http://arxiv.org/abs/hep-lat/0103029} {arXiv:hep-lat/0103029 [hep-lat]}
  \BibitemShut {NoStop}%
\bibitem [{\citenamefont {Follana}\ \emph {et~al.}(2007)\citenamefont
  {Follana}, \citenamefont {Mason}, \citenamefont {Davies}, \citenamefont
  {Hornbostel}, \citenamefont {Lepage}, \citenamefont {Shigemitsu},
  \citenamefont {Trottier},\ and\ \citenamefont {Wong}}]{Follana:2006rc}%
  \BibitemOpen
  \bibfield  {author} {\bibinfo {author} {\bibfnamefont {E.}~\bibnamefont
  {Follana}}, \bibinfo {author} {\bibfnamefont {Q.}~\bibnamefont {Mason}},
  \bibinfo {author} {\bibfnamefont {C.}~\bibnamefont {Davies}}, \bibinfo
  {author} {\bibfnamefont {K.}~\bibnamefont {Hornbostel}}, \bibinfo {author}
  {\bibfnamefont {G.~P.}\ \bibnamefont {Lepage}}, \bibinfo {author}
  {\bibfnamefont {J.}~\bibnamefont {Shigemitsu}}, \bibinfo {author}
  {\bibfnamefont {H.}~\bibnamefont {Trottier}}, \ and\ \bibinfo {author}
  {\bibfnamefont {K.}~\bibnamefont {Wong}} (\bibinfo {collaboration} {HPQCD,
  UKQCD}),\ }\href {\doibase 10.1103/PhysRevD.75.054502} {\bibfield  {journal}
  {\bibinfo  {journal} {Phys. Rev.}\ }\textbf {\bibinfo {volume} {D75}},\
  \bibinfo {pages} {054502} (\bibinfo {year} {2007})},\ \Eprint
  {http://arxiv.org/abs/hep-lat/0610092} {arXiv:hep-lat/0610092 [hep-lat]}
  \BibitemShut {NoStop}%
\bibitem [{\citenamefont {Bazavov}\ \emph {et~al.}(2013)\citenamefont {Bazavov}
  \emph {et~al.}}]{Bazavov:2012xda}%
  \BibitemOpen
  \bibfield  {author} {\bibinfo {author} {\bibfnamefont {A.}~\bibnamefont
  {Bazavov}} \emph {et~al.} (\bibinfo {collaboration} {MILC}),\ }\href
  {\doibase 10.1103/PhysRevD.87.054505} {\bibfield  {journal} {\bibinfo
  {journal} {Phys. Rev.}\ }\textbf {\bibinfo {volume} {D87}},\ \bibinfo {pages}
  {054505} (\bibinfo {year} {2013})},\ \Eprint {http://arxiv.org/abs/1212.4768}
  {arXiv:1212.4768 [hep-lat]} \BibitemShut {NoStop}%
\bibitem [{\citenamefont {Tanabashi}\ \emph {et~al.}(2018)\citenamefont
  {Tanabashi} \emph {et~al.}}]{Tanabashi:2018oca}%
  \BibitemOpen
  \bibfield  {author} {\bibinfo {author} {\bibfnamefont {M.}~\bibnamefont
  {Tanabashi}} \emph {et~al.} (\bibinfo {collaboration} {Particle Data
  Group}),\ }\href {\doibase 10.1103/PhysRevD.98.030001} {\bibfield  {journal}
  {\bibinfo  {journal} {Phys. Rev.}\ }\textbf {\bibinfo {volume} {D98}},\
  \bibinfo {pages} {030001} (\bibinfo {year} {2018})}\BibitemShut {NoStop}%
\bibitem [{\citenamefont {Edwards}\ and\ \citenamefont
  {Joo}(2005)}]{Edwards:2004sx}%
  \BibitemOpen
  \bibfield  {author} {\bibinfo {author} {\bibfnamefont {R.~G.}\ \bibnamefont
  {Edwards}}\ and\ \bibinfo {author} {\bibfnamefont {B.}~\bibnamefont {Joo}}
  (\bibinfo {collaboration} {SciDAC, LHPC, UKQCD}),\ }\bibfield  {booktitle}
  {\emph {\bibinfo {booktitle} {{Lattice field theory. Proceedings, 22nd
  International Symposium, Lattice 2004, Batavia, USA, June 21-26, 2004}}},\
  }\href {\doibase 10.1016/j.nuclphysbps.2004.11.254} {\bibfield  {journal}
  {\bibinfo  {journal} {Nucl. Phys. Proc. Suppl.}\ }\textbf {\bibinfo {volume}
  {140}},\ \bibinfo {pages} {832} (\bibinfo {year} {2005})},\ \bibinfo {note}
  {[,832(2004)]},\ \Eprint {http://arxiv.org/abs/hep-lat/0409003}
  {arXiv:hep-lat/0409003 [hep-lat]} \BibitemShut {NoStop}%
\end{thebibliography}
%

\end{document}